\documentclass[]{aa}

\usepackage{graphicx,times,psfig}
\usepackage{graphics}

\begin{document}
   \title{High resolution soft X-ray spectroscopy of the elliptical galaxy NGC~5044}
   \subtitle{Results from the reflection grating spectrometer on-board {\it XMM-Newton}}

   \author{
	T. Tamura \inst{1}
	\and
	J. S. Kaastra \inst{1}
         \and
	K. Makishima \inst{2}
         \and
	I. Takahashi \inst{2}
	}

	   \offprints{T.Tamura}
	\mail{T.Tamura@sron.nl}

\institute{SRON National Institute for Space Research, 
              Sorbonnelaan 2, 3584 CA Utrecht, The Nether\-lands 
	\and
	Department of Physics, The University of Tokyo,
		7-3-1 Hongo, Bunkyo-ku, Tokyo, 113-0033	
}
   \date{Received ?; accepted ?}

\abstract{
The results from an X-ray spectroscopic study of the giant elliptical galaxy NGC~5044 in the center of a galaxy group are presented.
The line dominated soft X-ray spectra (mainly Fe-L and \ion{O}{viii}~${\rm Ly\alpha}$) from the diffuse gas are resolved for the first time in this system 
with the Reflection Grating Spectrometers on-board {\it XMM-Newton}
and provide a strong constraint on the temperature structure.
The spectra integrated over 2\arcmin\ ($\sim$ 20~kpc) in full-width can be described 
by a two temperature plasma model of 0.7~keV and 1.1~keV.
Most of the latter component is consistent with originating from off-center regions.
Compared to the isobaric cooling flow prediction,
the observation shows a clear cut-off below a temperature of $0.6\pm0.1$~keV.
Furthermore, the Fe and O abundances within the central 10--20~kpc in radius are accurately measured to be $0.55\pm0.05$ and $0.25\pm0.1$ times the solar ratios, respectively.
The observed cut-off temperature of this galaxy and other central galaxies in galaxy groups and clusters are compared with the scale of the galaxy and properties of the surrounding intra-cluster medium.
Based on this comparison, the origin of the lack of predicted cool emission is discussed.
\keywords{Galaxies: individual: NGC 5044 --
Galaxies: clusters: general -- 
Galaxies: abundances --
X-rays: galaxies: clusters 
}}

\titlerunning{Soft X-ray spectroscopy of the elliptical galaxy NGC~5044}
\maketitle

\section{Introduction}
At the center of X-ray luminous galaxy clusters and groups, 
a giant elliptical galaxy is often found.
The formation of the galaxy most likely involves a series of galaxy mergers.
Around these regions, 
X-ray emission is also peaked, indicating a concentration of the hot gas (one to a few keV).
The central gas density ($>10^4$~m$^{-3}$) may result in gas cooling along with mass in-flow (cooling flow; CF; e.g. Fabian \cite{fabian94}),
although the predicted flow velocity has not been directly measured, and the fate of cooled gas has not been established.
Instead of in-flows, the cooling may be counterbalanced by heating mechanisms, such as thermal conduction (e.g. Takahara \& Takahara \cite{takahara79}), magnetic reconnection (e.g. Soker \& Sarazin \cite{soker90}; Makishima et al. \cite{makishima01}), or AGN activity (e.g. Churazov et al. \cite{churazov01}).
In any case, this cooling, mass in-flow, and heating processes should involve energy and/or mass transfers on the largest scale in the universe.

X-ray observations established the presence of plasma 
near the cluster center considerably cooler than that found further out in the cluster (e.g., Canizares et al. \cite{canizares79}).
However, a series of innovative X-ray results on nearby clusters have cast a serious doubt on the reality of CF hypothesis.
In particular, detailed spectroscopy with {\it ASCA}, {\it XMM-Newton}, and {\it Chandra} in representative clusters
revealed a lack of soft X-ray emission as predicted by a simple CF picture (e.g. Ikebe et al. \cite{ikebe99}; Peterson et al. \cite{peterson01}; David et al. \cite{david01}; but also see e.g. Schmidt et al. \cite{schmidt01}).
The detailed spatial and thermal structure of this cool component has not yet been systematically measured.
Consequently, the origin of the lack of cool emission is not fully understood.

Here we report the RGS spectroscopic observations of the core of the NGC~5044 galaxy group (WP23).
X-ray observations showed that the X-ray emitting gas is distributed over the group and centered on the giant elliptical galaxy, NGC~5044 (David et al. \cite{david94}; Fukazawa et al. \cite{fukazawa96}).
The gas temperature is about 1~keV with a central drop down to 0.8~keV.
Compared to these {\it ROSAT} and {\it ASCA} observations, the RGS provides much better spectral resolution in the 0.3--2.0~keV band; at 1~keV these instruments have energy resolution ($E/\Delta E$) of
$\sim2, \sim10, {\rm and} \sim200$, respectively.
This merit of the RGS overwhelms its two disadvantages for extended sources,
that it cannot yield two-dimensional distributions of the emission,
and that the source extent degrades the energy resolution.
We present here one of the best quality soft X-ray spectra of elliptical galaxies, 
and use the spectra to constrain the thermal structure and metallicity within the central 10-20~kpc in radius of NGC~5044.

Throughout this paper,
we assume the Hubble constant to be $H_0 = 100h$ km s$^{-1}$Mpc$^{-1}$
and use the 90\% confidence level.
At the redshift of 0.0090 (from the he NASA/IPAC Extragalactic Database; NED) for NGC~5044, 1\arcmin\ corresponds to $7.7h^{-1}$~kpc.
We use $4.68\times10^{-5}$ in number for the solar Fe abundance.

\section{Observations}
{\it XMM-Newton} observations of NGC~5044 were performed on 2001 January 12.
The RGS dispersion axis was oriented at a position angle of $114$ degrees (North to East).
We obtained a useful RGS exposure time of 22~ksec.
Detailed descriptions of the {\it XMM-Newton} satellite and the RGS instrument are found in Jansen et al. (\cite{jansen01}) and den Herder et al. (\cite{herder01}), respectively.
Buote et al. (\cite{buote02}) analyzed the EPIC data of this observation and submitted a paper on the spatial properties of the X-ray emitting gas.

\section{Analysis and Results}
\subsection{Data extraction and fitting method}
We analyzed the RGS data in a way similar to previous reports (e.g. Tamura et al. \cite{tamura01b}), 
but with improved calibration of the instrumental responses and background.
For basic data processing, we used the Science Analysis System (SAS ver.5.3.3).
We extracted two sets of spectra from two different regions,
$|w|<1'$ and $1'<|w|<2'$, where $w$ is the distance from the X-ray maximum in the cross-dispersion direction.
Hereafter we refer to these spectra as W02 and W24, respectively. 
Because of the principle of the instrument, 
each spectrum contains emission from a range of different projected positions.
Based on the observed EPIC/MOS image in the RGS energy band, 
we estimate this mixing, as given in Table~\ref{tbl:rgs-mix}.
This estimate implies that more than 80\% of the flux originates from 
$r<2'$ and $1'<r<4'$ for W02 and W24, respectively, where $r$ is the projected radius from the center.
Since the instrumental spatial resolution, $\sim 15$\arcsec\ in FWHM, is smaller than our extraction widths, 
the PSF effects insignificantly on this estimation.

\begin{figure}[h]
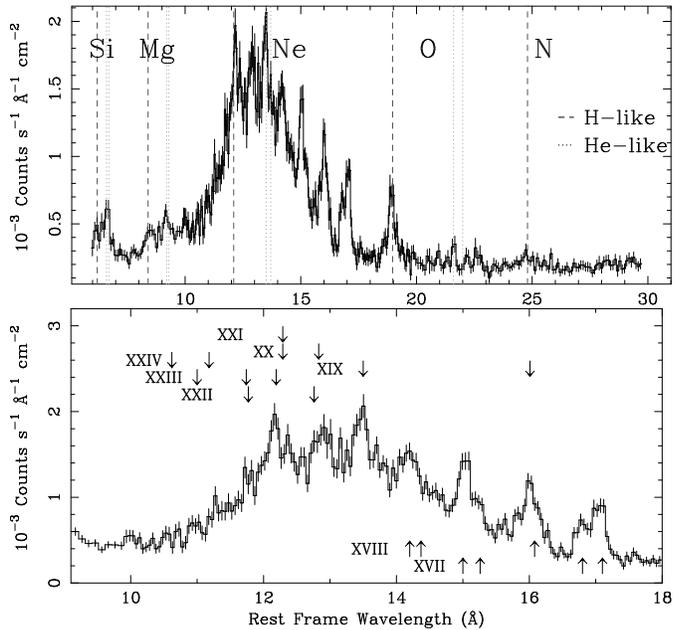

	\resizebox{\hsize}{!}{\includegraphics[angle=-90]{h3947f1a.ps}}
	\resizebox{\hsize}{!}{\includegraphics[angle=-90]{h3947f1b.ps}}
      \caption[]{The RGS spectrum extracted from the central 4\arcmin\ in full-width from the NGC~5044 group. 
	The spectrum is corrected for redshift and effective area, but not for the background.
	Four spectra from the two instruments and two spectral orders (m=-1 and -2) are combined.
The bottom panel is a zoom of a part of the spectra. The positions of H- and He-like ions (top panel) and Fe-L transitions (bottom) are indicated.}
         \label{fig:raw-spe}
   \end{figure}

We present the spectra (sum of W02 and W24) in Fig.~\ref{fig:raw-spe}.
Thus, we can resolve lines from H- and/or He-like ions of Si, Mg, and O, 
along with a series of Fe-L lines.
The observed Fe-L lines are from ionization states ranging from \ion{Fe}{xvii} to \ion{Fe}{xxi}, 
suggesting emission from their maximum formation temperatures ranging from 0.4~keV to 1.1~keV.

\begin{table}
\begin{center}
\caption[]{The estimated flux mixing effect, based on the MOS1 image$^{\mathrm{a}}$.}
\label{tbl:rgs-mix}
\begin{tabular}{lcc}
\hline
2D-radius (\arcmin)	& W02	& W24 \\
\hline
$<$1			& 0.28	& 0\\
1--2			& 0.16	& 0.15\\
2--4			& 0.07	& 0.10\\
4--5			& $<0.03$  & $<0.04$ \\
\hline
\end{tabular}
\begin{list}{}{}
\item[$^{\mathrm{a}}$] The numbers give fractions of fluxes contributed by individual two-dimensional annuli to the RGS data integration regions, relative to the total flux within 5\arcmin\ in radius.
\end{list}
\end{center}
\end{table}

The emission is extended on larger scales than the instrumental spatial resolution 
resulting in degradation of the line spread function (LSF).
We approximated this effect by
convolving the source surface brightness profile with the RGS response for a point source (Kaastra et al. \cite{kaastra02a}).
In this way, the line width approximately becomes 
\begin{equation}
[l^2+(0.12~{\mathrm \AA} \cdot\theta/1\arcmin)^2]^{0.5} 
\end{equation}
, where $l$ and $\theta$ are intrinsic instrumental line width and source spatial width, respectively.
Here we modeled the brightness based on the observed MOS image.
Due to the mirror PSF, the MOS image is more extended than the actual brightness profile.
To compensate for this for the W02 spectral fitting, we re-scaled the brightness distribution to match the observed RGS profiles of the \ion{O}{viii} and Fe-L lines, observed in the 14--20.5~\AA\ range.
We found a scaling factor in wavelength width of $f=0.83$, which is used for the spectral fitting below. 
This corresponds to $\theta = $ (observed width with MOS) $ \times f $.
For W24, which has too poor statistics to determine the scaling factor, 
we simply used $f=1$.

We limited the analysis to the wavelength band of 8--28~\AA, 
where the estimated background is typically less than 50\% of the source flux.
The background spectrum and its variability were estimated from several blank-sky field observations ($\sim 400$~ks in total exposure).
A 30\% systematic error was assigned to the background counts (den Herder et al. \cite{herder02}).
The first order spectra from the RGS1 and RGS2 were fitted with the same model simultaneously.

\subsection{Isothermal model fitting}
To characterize the observed spectra and find the temperature and metallicity,
we fitted the spectra with a collisonal ionization equilibrium model (CIE; implemented in the SPEX package; Kaastra et al. \cite{kaastra02a}) modified by photoelectric absorption and cosmological redshift (1T model). 
The absorbing column density was fixed to the Galactic value, $4.9\times 10^{24}$~m$^{-2}$, obtained from the \ion{H}{i} map (Dickey \& Lockman \cite{dickey90} using NASA's W3nH tool).
The abundance ratios relative to the solar value of He was fixed to unity (He/H=1), 
while those of other heavy elements, except for O, were fixed to that of Fe (e.g., Si/Fe=1).
The solar abundances are taken from Anders and Grevesse (\cite{anders}).
The free parameters are the emission measure ($EM$), the temperature, Fe/H, and O/Fe ratios.

\begin{figure}
	\resizebox{\hsize}{!}{\includegraphics[angle=-90]{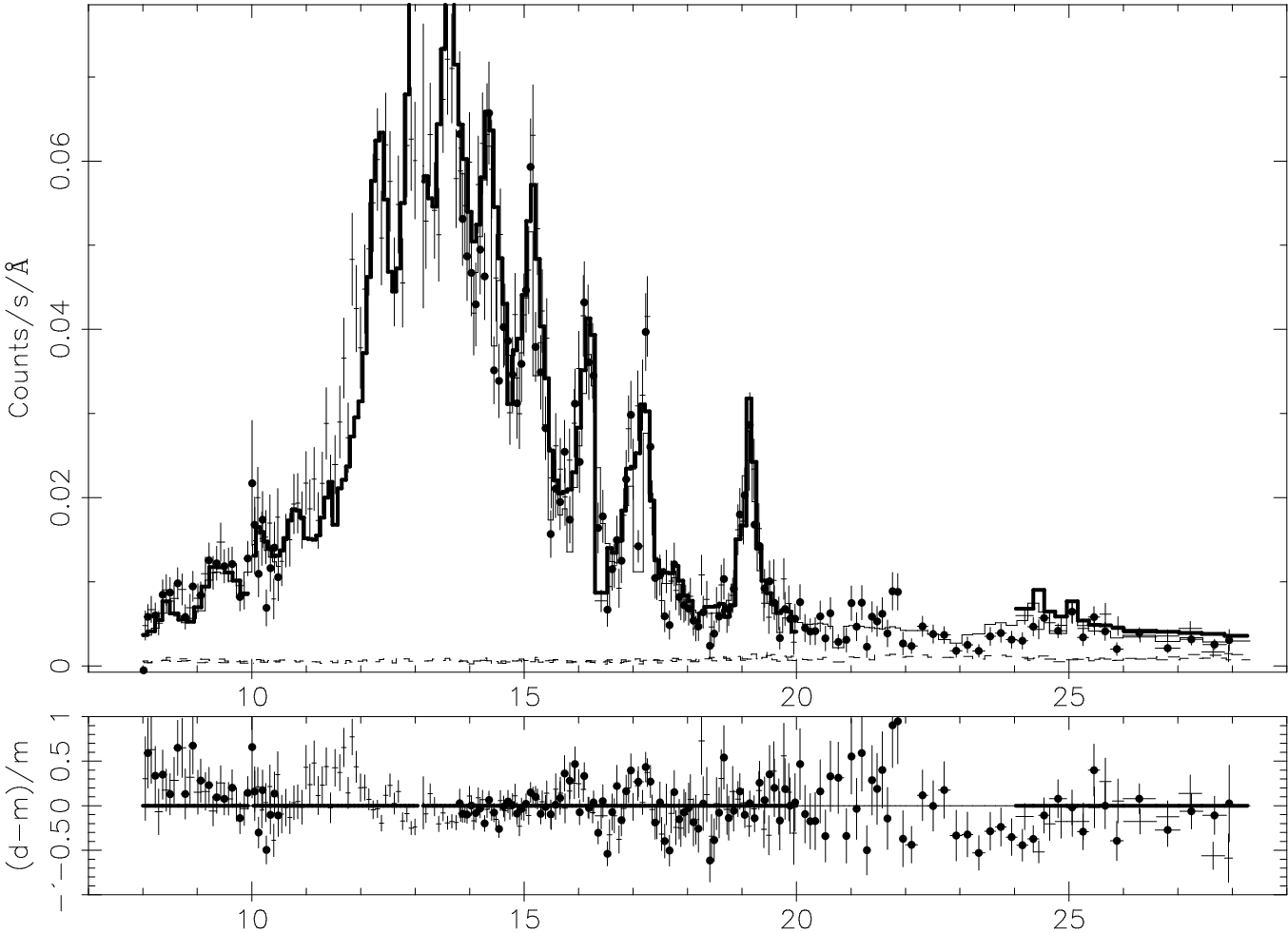}}
	\resizebox{\hsize}{!}{\includegraphics[angle=-90]{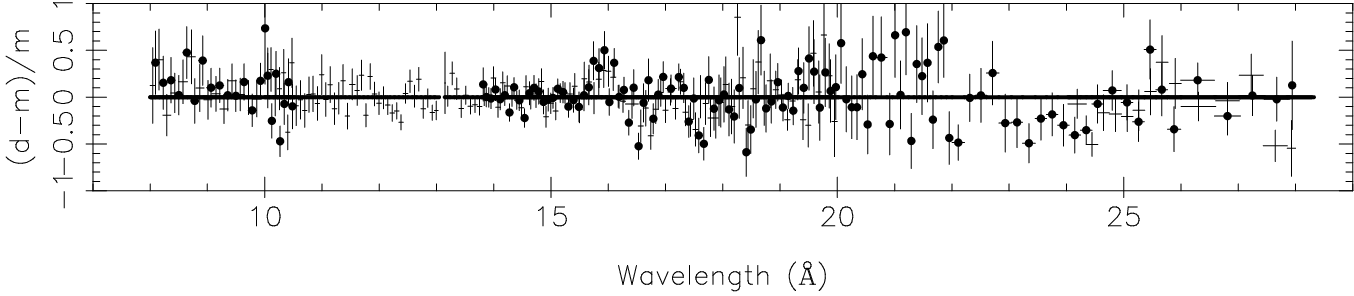}}
      \caption[]{The RGS first order spectra extracted from a full-width of 2\arcmin\ (W02; filled-circle for RGS1, cross for RGS2). Top, middle, and bottom panels show the data compared with the best-fit 1T model (histogram), 
the residuals (data-model/model) for the 1T model, and those for the multi-T model, respectively. In the top panel, the background spectra are shown with dashed-line histograms.}
         \label{fig:rgs-fit-w2}
	\resizebox{\hsize}{!}{\includegraphics[angle=-90]{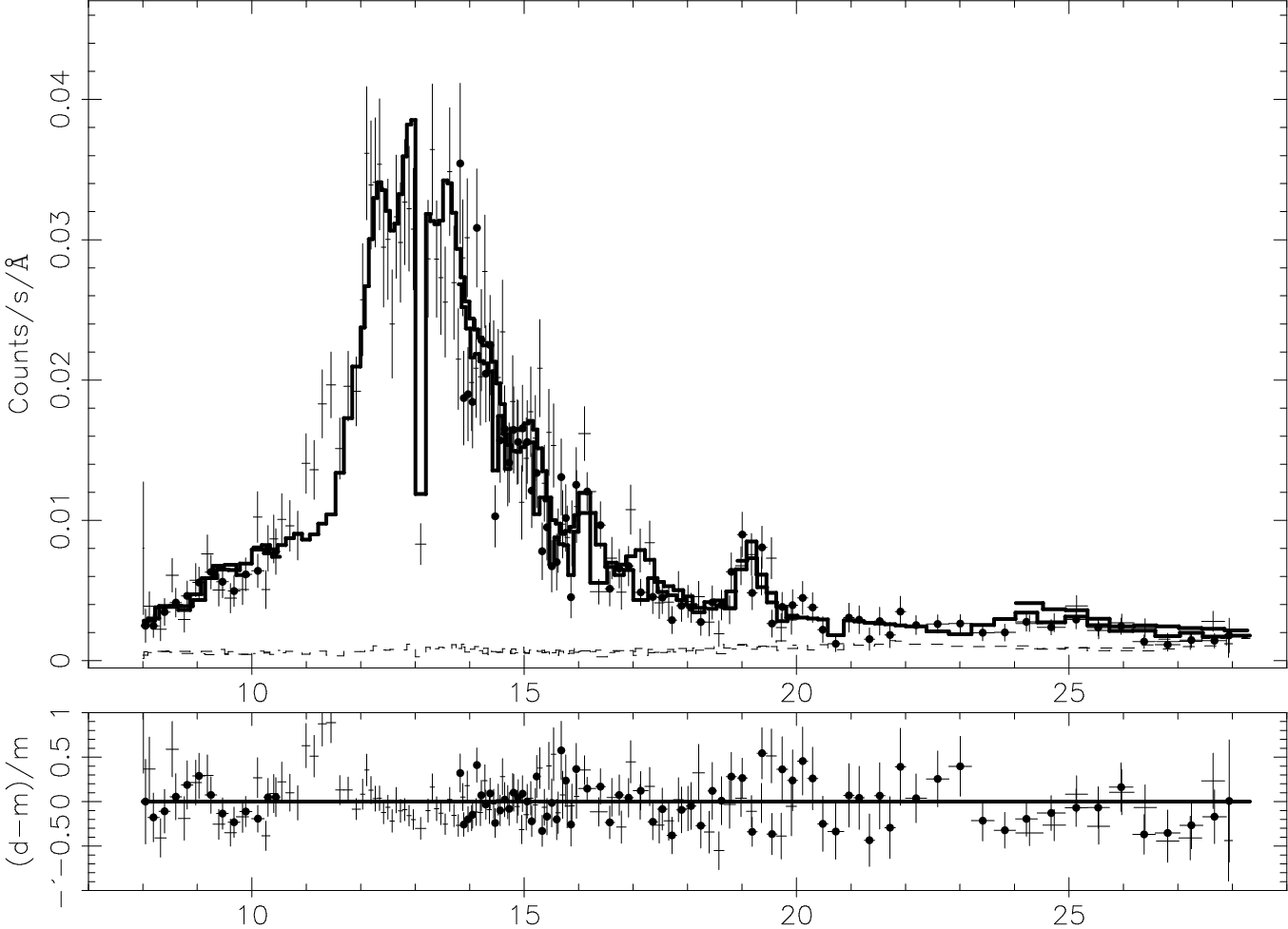}}
	\resizebox{\hsize}{!}{\includegraphics[angle=-90]{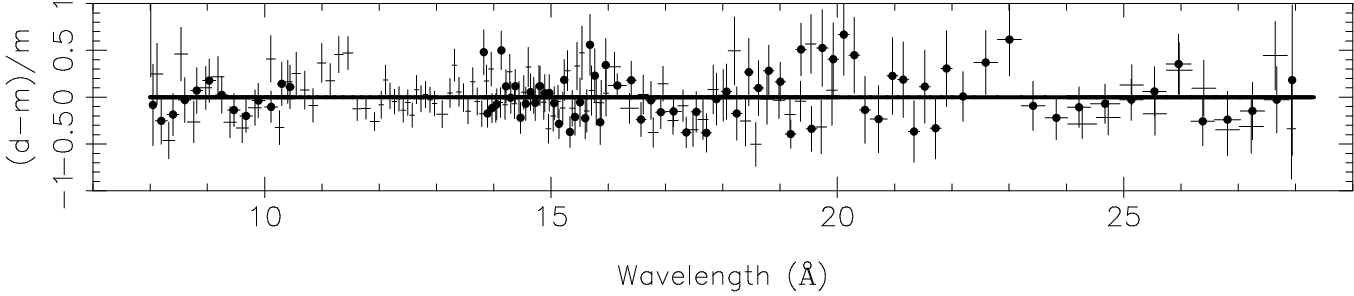}}
      \caption[]{Same as the previous figure, but for the W24 spectra.}
         \label{fig:rgs-fit-w2-4}
   \end{figure}

The results of fitting with the 1T model are shown in Table~\ref{tbl:rgs-fits} and Figures~\ref{fig:rgs-fit-w2}-\ref{fig:rgs-fit-w2-4}.
Although the fits are not acceptable, the 1T model reproduces the global spectral structure fairly well,
indicating that the emission is dominated by a single temperature component.
The representative temperatures derived from the 1T fit, 0.8~keV (W02) and 0.9~keV (W24), are consistent with the previous ROSAT and ASCA measurements (David et al. \cite{david94}; Fukazawa et al. \cite{fukazawa96}).

Significant residuals are found around 11--12~\AA,
where lines from \ion{Fe}{xxii} -- \ion{Fe}{xxiv} are expected, 
suggesting the presence of emission hotter than the effective temperature.
Another excess can be seen around 16~\AA\ in both the W02 and W24 spectra;
the model underestimates the data. 
This is unlikely due to the incorrect modeling of the LSF, 
because the model reproduces other lines, which should have similar brightness distribution and hence similar LSF.
This is perhaps intrinsic to the emission model of lines (mainly \ion{Fe}{xviii} and \ion{O}{viii}~Ly$_{\beta}$).
In fact, the RGS spectrum of Capella also shows excess around this band above the SPEX/CIE model (Audard et al. \cite{audard01}).
Another residual above the 1T model can be seen around 22~\AA. 
Although this is close to the red-shifted position of \ion{O}{vii} lines, 
we presume that this is due to residual instrumental hot pixels/columns instead of emission from the source, 
because it is too narrow compared to other lines.
There are some hot pixels/columns in the RGS CCDs whose duty cycles are too low to detect statistically within normal integration time.

\begin{table*}
\begin{center}
\caption[]{The results of model fits to the RGS spectra.}
\label{tbl:rgs-fits}
\begin{tabular}{lccccccc}
\hline 
(1)	& (2) $EM$ & (3) T& (4) Fe/H & (5) O/Fe & (6) $\Delta N_{\mathrm H}$ & (7) $\chi^2/\nu$  & model\\
\hline
W02$^{\mathrm{a}}$	& 3.8	& 0.78	& 0.41	& 0.54	& 0f & 478/278	& 1T\\ 
W24$^{\mathrm{a}}$ & 2.0  & 0.88	& 0.38  & 0.43  & 0f & 284/183  & 1T \\ 
\hline
W02	& -	& 1.07$\pm0.03$& 0.55$\pm 0.05$& 0.51$\pm 0.06$	& 0f & 339/273	& multi-T\\
W24 	& -	& 1.07f  & 0.53$\pm 0.05$	  & 0.42$\pm 0.12$      & 0f & 242/179 & multi-T\\
\hline
W02	& -	& 1.07f	& 0.48$\pm 0.05$ & 0.57$\pm 0.06$& 7.5$^{+5.0}_{-1.0}$ & 353/279	& ICF+$\Delta N_{\mathrm H}$\\ 
\hline
\end{tabular}
\begin{list}{}{}
\item[(1)] The spectral extraction region; W02 and W24 as described in the text.
\item[(2)] The volume emission measure ($h^{-2}10^{70}$m$^{-3}$).
\item[(3)] The temperature (keV). The highest temperature ($T_0$) in the case of a multi-T and ICF models.
\item[(6)] Column density ($10^{25}$~m$^{-2}$). The covering factor of the excess absorption for the hot component is fixed to be 0.1.
\item[$^{\mathrm{a}}$] No errors are given for these fits, because the models are far from acceptable.
\item[f] Fixed parameters.
\end{list}
\end{center}
\end{table*}

\subsection{Multi temperature model fitting}
Given the 1T fit results, 
we attempt to obtain a better fit by introducing a combination of multiple CIE components with different temperatures.

In the model fitting,
we fixed the temperature separation as $T_{i}/T_{i+1}\equiv F=\mathrm{constant}$, where $T_{i}$ is the $i$-th temperature.
A smaller $F$ gives a better description of true $EM$ structure, 
but at the same time it causes coupling between the components.
Due to the nature of the Fe-L line emissivities, 
we cannot resolve the $EM$ structure with resolution finer than about $F=1.25$, 
regardless of the data quality (Kaastra et al. \cite{kaastra02b}).
Given the present spectral quality, we use $F\simeq 1.5$.
We allowed each component to have a free normalization, $EM(T_i)$, 
but fixed abundances among components.
Therefore, additional free parameters are $T_0$ and $EM(T_i)$, where $T_0$ is the temperature of the hottest component.
We limited $T_{i}$ to be hotter than 0.1~keV since the RGS spectrum is not sensitive to much cooler emission.
We used the best-fit Fe and O abundances obtained in the 1T fit as initial parameters. 
Furthermore, for W02 we started with $T_0 = 1.2$~keV, which is 1.5 times the representative temperature.
For the W24 fit, we fixed $T_0$ to the best-fit value of the W02 fit (1.07~keV) ,
because we are interested in the difference in the $EM$ shape between W02 and W24, 
instead of the value of $T_0$.

\begin{figure}
\resizebox{\hsize}{!}{\includegraphics[angle=-90]{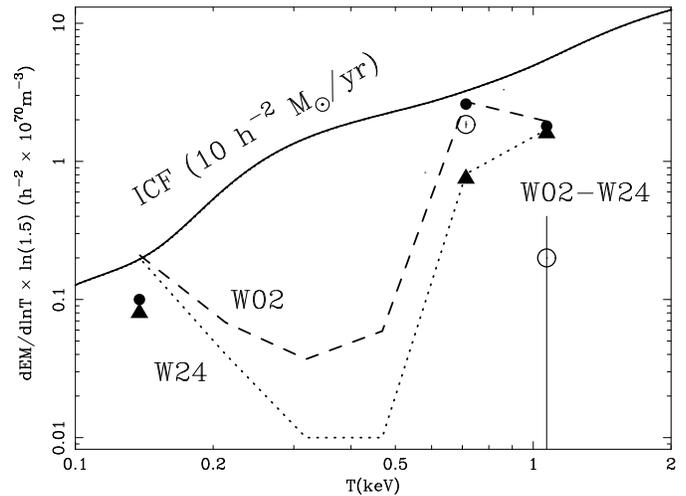}}
      \caption[]{The constraint on the differential emission measure in terms of $\frac{\rm{d}EM}{\rm{d}\ln T}$. 
The filled circles and triangles indicate the best-fit values for the W02 and W24 spectra, respectively, 
while the open circles indicate the estimated values for the emission from the galaxy core ($r<1'$) only of the 0.7~keV and 1.1~keV components.
The dashed and dotted lines indicate upper limits on values  for W02 and W24, respectively.
The model predicted from the isobaric cooling flow with a mass deposition rate of 10~$h^{-2}$~\hbox{M$_{\odot}$}yr$^{-1}$ is shown as the solid line. 
}
         \label{fig:em}
\end{figure}

As shown in Table~\ref{tbl:rgs-fits} and Figures~\ref{fig:rgs-fit-w2}-\ref{fig:rgs-fit-w2-4}, 
the fits improved significantly ($\Delta \chi^2 = $139 and 42 for W02 and W24, respectively).
The W02 spectra can be described by the sum of two CIE components ($T_0 \sim $ 1.1~keV and $T_1 \sim $ 0.7~keV).
A relatively small contribution from a 0.14~keV component is marginally suggested.
No other component (0.2--0.6~keV) is required as illustrated in Fig.~\ref{fig:em}, 
where we compare the derived $EM$ distribution with that predicted from an isobaric cooling flow model (ICF; Canizares et al. \cite{canizares88}).
The ICF model is the simplest description of emission from a cooling gas without any heating; 
the differential $EM$ (${\mathrm{d}EM}/{\mathrm{d}T}$) is proportional to the inverse of the cooling function and the mass deposition rate.
This comparison clearly indicates a lack of the emission from 0.2--0.6~keV gas, 
relative to the emission from 0.7~keV gas, 
contrary to the prediction of the ICF model.

A direct comparison of the observed spectrum with the ICF model shows 
that the 0.7~keV component nicely reproduces the fluxes and ratios of major resolved lines (\ion{Fe}{xvii}, \ion{Fe}{xviii}, and \ion{O}{viii}) at least within a wavelength range of 14--20~\AA.
Further, the 0.5~keV or 0.3~keV components of the ICF model clearly overpredict line fluxes of \ion{Fe}{xvii} at 15.0~\AA\ and 17.0~\AA.
Finally, in the 0.2~keV and 0.1~keV ICF model components, 
the line fluxes from \ion{O}{viii} and \ion{O}{vii} become 
  so high relative to the Fe-L lines that they are inconsistent with the data.

In addition to O and Fe, 
we attempted to constrain the abundances of N, Ne, Mg, and Si.
The H- and/or He-like lines of these elements are covered by the RGS.
To derive individual abundances separately, 
we fixed all parameters except the relevant abundance 
and fitted within the limited spectral wavelength bands around the expected line from each element (Table~\ref{tbl:rgs-fits-abun}).
We used the best-fit multi-temperature model.
Table~\ref{tbl:rgs-fits-abun} shows the result for W02.
Among these abundances, those of N and Si could have systematic errors larger than the statistical errors shown in the Table.
In the case of N, emission from gas cooler ($<$0.1~keV) than the RGS sensitivity may contribute to the line emission. 
The calibration of the RGS response around the Si line ($\sim 7$~\AA) is relatively less certain; 
  the Si abundance may be subject to systematic errors up to 20\%, which is similar to the statistical ones.
Therefore we conservatively conclude that these elemental abundances relative to Fe are consistent with the solar ratio within 50\% uncertainty.

\begin{table}
\begin{center}
\caption[]{The abundances of lighter elements relative to iron, in solar units$^{\mathrm{a}}$.}
\label{tbl:rgs-fits-abun}
\begin{tabular}{lcc}
\hline 
	& Abundance	& band$^{\mathrm{b}}$(\AA)\\
\hline
N/Fe	& $0.62\pm 0.5$ & 24-25 \\
Ne/Fe	& $0.55\pm 0.3$ & 11--14 \\
Mg/Fe	& $1.26\pm 0.2$ & 8--10 \\
Si/Fe	& $1.50\pm 0.3$ & 6--7.5 \\
\hline
\end{tabular}
\begin{list}{}{}
\item[$^{\mathrm{a}}$] Derived from the W02 spectra, using the best-fit multi-temperature model (Fe/H=0.55).
\item[$^{\mathrm{b}}$] The wavelength band used in the analysis.
\end{list}
\end{center}
\end{table}

\subsection{Absorption by cold material}
One possibility to hide the cool emission is absorption by blobs of cold material.
If the cold material is located close to or associated to the X-ray emitting cool component, the expected X-ray emission will be selectively absorbed. 
This cold material should also absorb a part of the hot component emission. 
In the ICF model, the relative volume occupied by a temperature component is proportional to 
${dT}/\left[\Lambda(T) n^2\right] \propto {T^2dT}/{\Lambda(T)}$, where $\Lambda(T)$ and $n$ are the emissivity and the gas density.
This implies that the volume occupied by the 0.4~keV component is roughly 10\% of the volume of the 1~keV component.
Therefore, the covering fraction for the hot component to be absorbed is expected to be also $\sim 10\%$.
To examine this possibility, we modeled the RGS spectrum by the ICF model and differential absorption. 
Assuming that 100\% and 10\% of the cool ($0.1<k_\mathrm{B}T/\mathrm{keV} < 0.6$) and hot ($0.7<k_\mathrm{B}T/\mathrm{keV} < 1.1$) emission are absorbed by the same column density, 
we have confirmed that a column density higher than $6.5\times 10^{25}$~m$^{-2}$ is required to describe the observed spectrum (Table~\ref{tbl:rgs-fits}).  
Note that the best-fit Fe and O abundances did not change significantly in this modeling. 

\begin{figure}[h]
\resizebox{0.8\hsize}{!}{\includegraphics[]{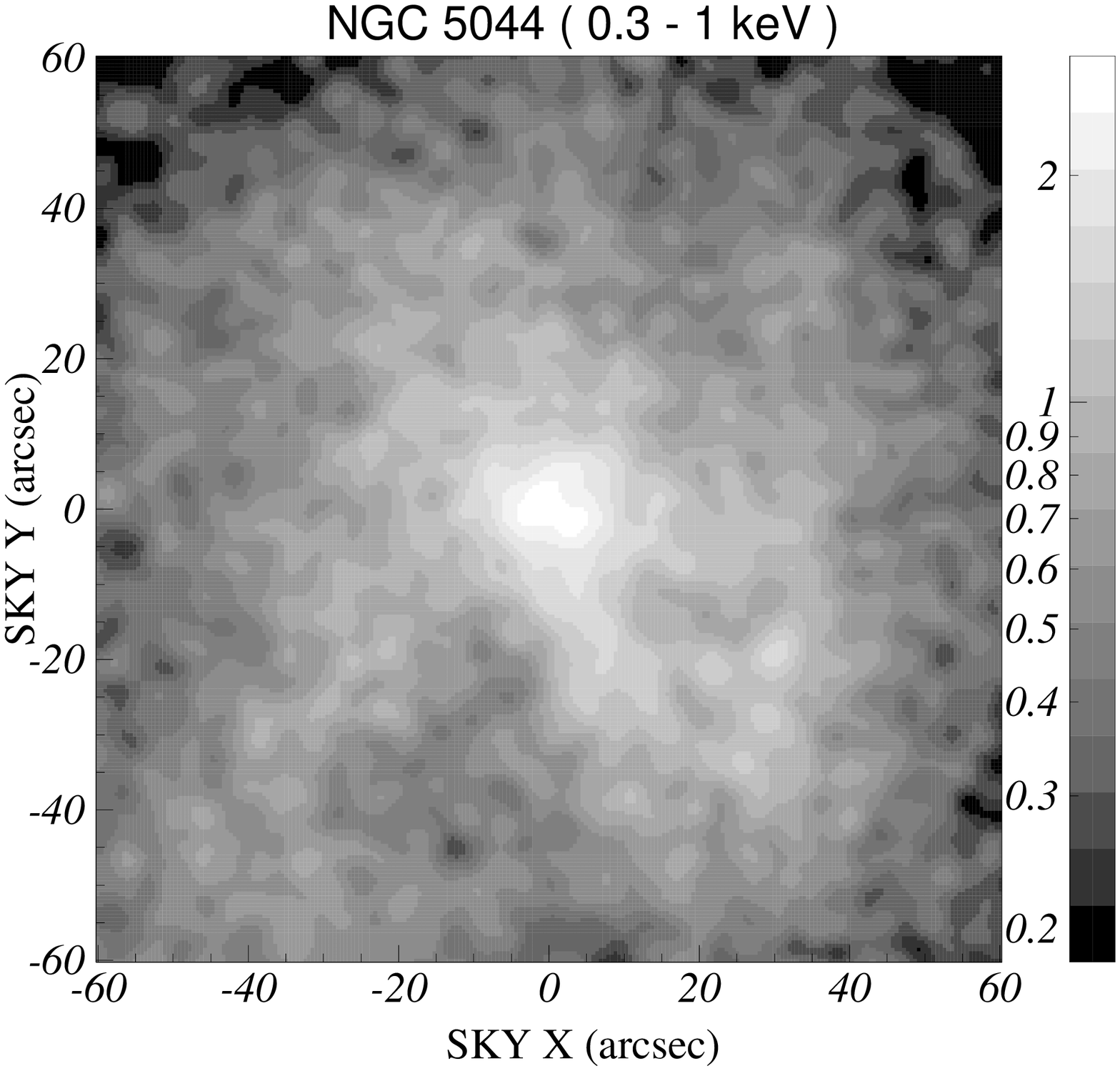}}
\resizebox{0.8\hsize}{!}{\includegraphics[]{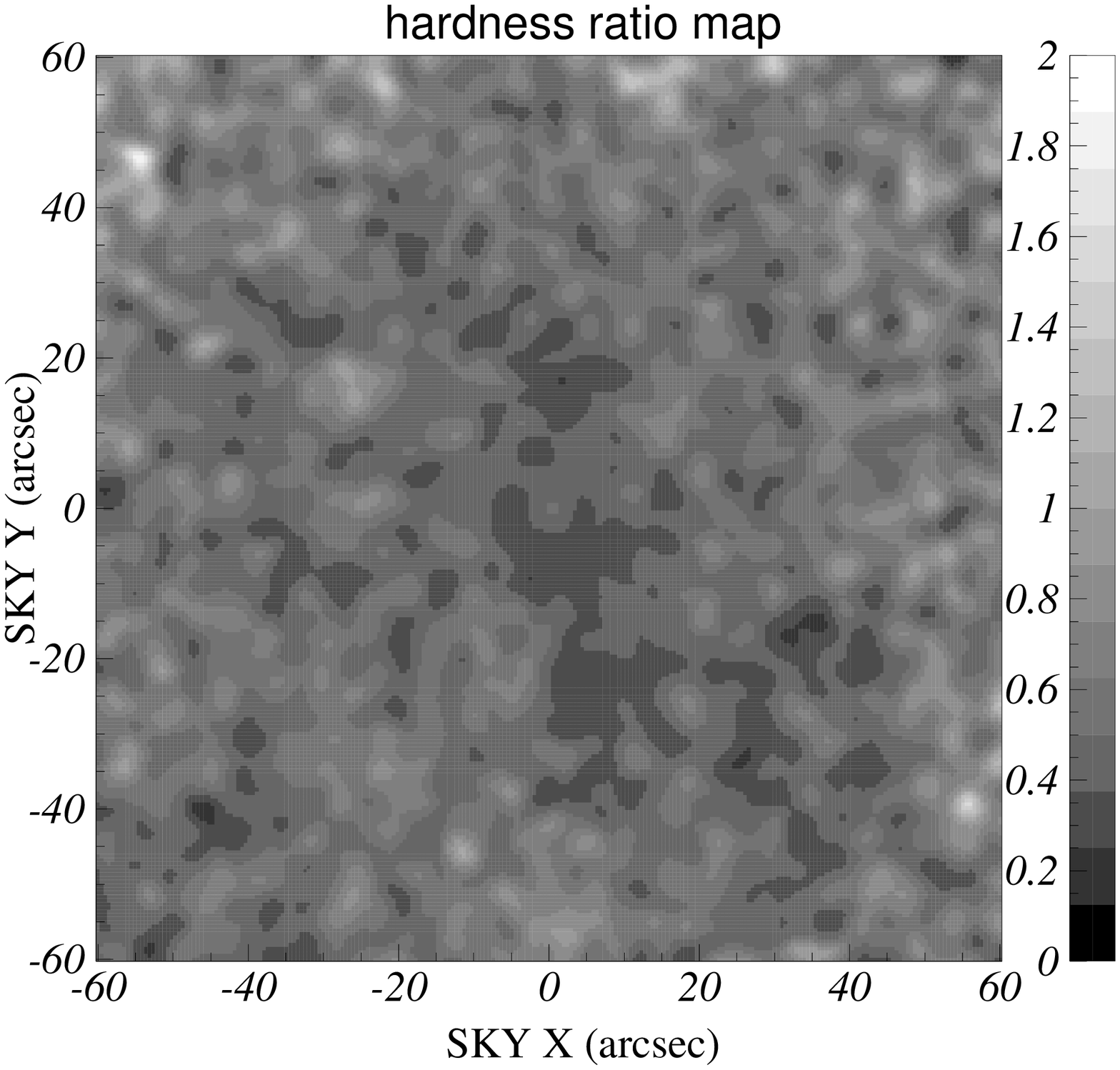}}
      \caption[]{{\it Chandra} ACIS-S image of the center of NGC~5044.
Top is the 0.3--1.0 keV brightness map smoothed with a Gaussian of $\sigma = 1''.5$, 
and bottom is a hardness ratio map (1--2~keV / 0.3--1~keV) of the same region. 
The scales are arbitrary in both panels.}
         \label{fig:chandra}
\end{figure}

If there is cold material with $N_{\mathrm H} \simeq 10^{26}$~m$^{-2}$,
more than 99~\% of the soft X-rays ($ < 0.7$~keV) must be absorbed, 
resulting in emission holes for external observers.
To examine whether there exist such brightness depressions,
we have analyzed archival {\it Chandra} data of NGC~5044.
As shown in Fig.~\ref{fig:chandra},
  there is no such effect near the center, 
nor drastic inhomogeneity of the emission hardness, down to a spatial scale of $\sim 10''$ ($\sim 2$~kpc).
Therefore, we conclude that this scenario is very unlikely.

\section{Discussion}
\subsection{Summary of the results}
We have successfully resolved a number of emission lines from the central regions of the NGC~5044 group.
As a result, we have found that the spectra extracted with a full-width of 2\arcmin\ (W02) 
can be described by a combination of a 0.7~keV and 1.1~keV CIE component.
Any cool component with temperature in the range of 0.2--0.5~keV and 0.1--0.2~keV is less than a few and ten percent of the overall emission measure ($EM$), respectively.

\subsection{Spatial temperature distribution}
The W02 spectra contain emission from different projected radii.
We can approximately subtract the contribution from the outer region 
by assuming that the spectral shape is azimuthally uniform and using 
the surface brightness distribution observed with the MOS.
The estimate in Table~\ref{tbl:rgs-mix} implies that
similar fluxes from a region of $1'<r<4'$, where $r$ is the projected radius,
are contaminated into the W02 and W24 spectra.
Therefore, by subtracting the W24 model from that of W02 
we can estimate the spectral nature of the galaxy core ($r<1$\arcmin ).
As shown by open circles in Fig.~\ref{fig:em}, 
the projected core (and hence the three-dimensional core) is dominated by the 0.7~keV component and the $EM$ of the 1.1~keV component is less than 20\% of that of the 0.7~keV emission.

This RGS result is consistent with the EPIC results by Buote et al. (\cite{buote02}).
They deprojected the observed spectra into three dimensional space and found that within and beyond $1'$ the temperature is $0.75\pm 0.1$~keV and $1.2\pm 0.1$~keV, respectively.

Because the separation between the two temperatures derived from the RGS (0.7~keV and 1.1~keV) is small, 
it is difficult to distinguish 
between (1) emission from two temperature components and 
(2) a continuous distribution in temperature between the two temperatures.  
The similarity in the obtained temperature values between the RGS and EPIC, 
which have different energy responses, 
suggests that model (1) is more likely than model (2). 

\subsection{Possibilities of resonance line scattering}
Resonance line scattering could affect line strengths from the core regions of X-ray bright clusters and elliptical galaxies (Gil'fanov et al. \cite{gilfanov}; Shigeyama \cite{shigeyama98}).
For the temperature of the core of NGC~5044, $\sim$ 0.7~keV, 
the highest optical depth is expected for \ion{Fe}{xvii} at 15.0~\AA\ (oscillator strength $f=2.7$).
In fact, Xu et al. (\cite{xu01}) detected resonant scattering effects in the 15.0~\AA\ line,  
from the RGS observations of the core of the elliptical galaxy NGC~4636. 
This galaxy has a temperature structure similar to that of NGC~5044.

In the case of NGC~5044, the optical depth of this line towards the galaxy center is 
estimated to be about 4 when there is no line broading due to bulk motion of the gas.
Then, the 15.0~\AA\ line photons would be scattered out of the core region,
resulting in a depression in W02 and an enhancement in W24.
However, there is no evidence for this in the observed spectra (Figures~\ref{fig:rgs-fit-w2}-\ref{fig:rgs-fit-w2-4}).
This is probably because we have integrated all emission within a full-width of 2\arcmin, 
which is wider than the emission {\em core} of this object, $\sim 30$\arcsec\ in radius.
Thus, we estimate the resonance scattering effects to be small.

\subsection{Origin of the lack of cool component}
According to our results and the above argument, 
  we presume that there is a cut-off in temperature, $T_{\rm cut}$, in the true $EM$ distribution.
In cores of medium-size to rich clusters, 
a similar lack of cool emission has been reported based on the high resolution RGS spectra (Kaastra et al. \cite{kaastra01}; Peterson et al. \cite{peterson01}; Tamura et al. 2001a, 2001b; Sakelliou et al. \cite{sakelliou02}; Xu et al. \cite{xu01}), as well as from {\it ASCA} (e.g. Makishima et al. \cite{makishima01}) and {\it Chandra} (e.g. David et al. \cite{david01}).
Among these, our result along with that of NGC~4636 by Xu et al.(\cite{xu01}) is unique because 
1) the quality of the observed spectra is among the best,
2) the observed $T_{\rm cut}$ value is significantly lower than for other sources studied,
and
3) the galaxy is surrounded by a relatively cooler (and lower pressure) intracluster medium (ICM) than in other cases.
A comparison with the ICF prediction suggests that some mechanism has heated or mixed the gas   with a total energy ($\sim 10^{35}$~W in 10~Gyr) and efficiency high enough to counterbalance the observed X-ray radiative cooling in this system.
Several ideas for the heating and mixing have been discussed (e.g. Makishima et al. \cite{makishima01}; Peterson et al. \cite{peterson01}; Fabian et al. \cite{fabian01}; Brighenti \& Mathews \cite{brighenti02}; B{\" o}hringer et al. \cite{hans02}).

\begin{figure}[h]
	\resizebox{\hsize}{!}{\includegraphics[]{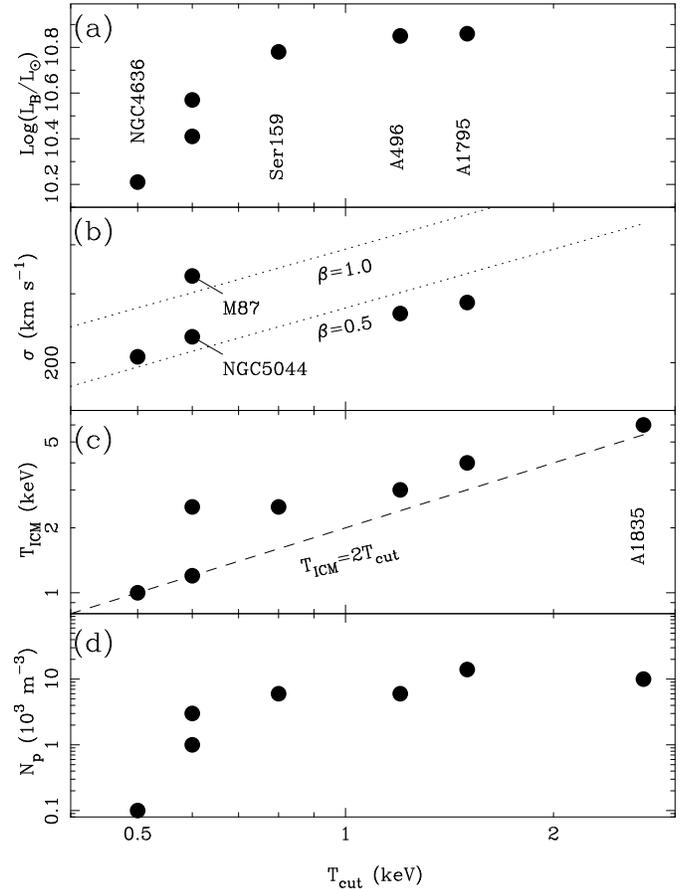}}
      \caption[]{The cut-off temperature ($T_{\rm cut}$) vs. {\bf (a)} optical blue luminosity and {\bf (b)} stellar velocity dispersion of the central galaxy, {\bf (c)} the temperature and {\bf (d)} the proton density of the ICM at radius of about 50$h^{-1}$~kpc, respectively, from top to bottom. 
The dotted lines in panel (b) correspond to $\frac{\mu m_{\rm p}\sigma^2}{kT_{\rm cut}} \equiv \beta = 0.5~{\rm and} 1.0$.
Dashed line in panel (c) corresponds to $T_{\rm ICM} = 2T_{\rm cut}$ (Ikebe \cite{ikebe01}). 
Data are taken from the NED, McElroy (\cite{mcelroy95}), Matsumoto et al. (\cite{matsumoto96}), Kaastra et al. (\cite{kaastra01}), Trinchieri et al. (\cite{trinchieri94}), Tamura et al. 2001a, 2001b, Peterson et al. (\cite{peterson01}), and Schmidt et al. (\cite{schmidt01}).
There is no blue luminosity nor velocity dispersion available for A~1835, 
perhaps because of the relatively large redshift of this target ($z=0.235$).
}
         \label{fig:comp}
   \end{figure}

To further understand the possible heating or mixing in these systems,
we compare the values of $T_{\rm cut}$ observed from several giant elliptical galaxies,
against other observables as shown in Fig.~\ref{fig:comp}.
These galaxies are in the outskirts (NGC~4636) or center (M~87) of the Virgo cluster, 
in the center of a group (NGC~5044), 
and in the centers of poor (S\'ersic 159-03) to rich  (A~496, A~1795 and A~1835) clusters. 
The value of $T_{\rm cut}$ was determined for all cases from the RGS spectra, 
but not uniformly analyzed 
and therefore include a systematic uncertainty up to 30\%.
As is clear from the figure,
the values of $T_{\rm cut}$ of these galaxies appear to be correlated
to their optical luminosity and the stellar velocity dispersion ($\sigma$),
as well as to the properties of the surrounding ICM.
Note that a correlation between the the  emission-weighted gas temperature and $\sigma$ has been reported for elliptical galaxies (e.g. Matsumoto et al. \cite{matsumoto97}; Matsushita \cite{matsushita01}): 
$\mu m_{\mathrm p}\sigma^2/kT = 0.5-1.0$, where $\mu \sim 0.6$ and $m_{\mathrm{p}}$ are the mean molecular weight and proton mass.
Furthermore, by fitting the {\it ASCA} spectra of $\sim 10$ cluster centers by a two temperature model,
Ikebe (\cite{ikebe01}) found a tight correlation between the two temperatures, 
which resembles the present correlation between $T_{\rm cut}$ and the ICM temperature (Fig.~\ref{fig:comp}c).
It is unclear, however, whether these correlations are intrinsic or due to the correlation of $T_{\rm cut}$ with the ISM temperature combined with the correlation of the ISM temperature with these parameters.

Based on the correlation between $T_{\rm cut}$ and the scale of the galaxy, 
at least a substantial part of the heating energy may originate from the galactic gravitational potential.
Possible energy-transfer mechanisms include 
an adiabatic compression along with the gas inflow (e.g. David et al. \cite{david94}) and
stellar motion through mass-loss or amplification of interstellar magnetic fields (e.g. Ikebe et al. \cite{ikebe99}).
These heating effects should be taken into account in relatively smaller systems like NGC~5044, where the gravitational energy is comparable to the thermal energy of the gas.
For the larger systems, however, the gravitational energy may be not enough to balance the large X-ray luminosity.
Alternatively,
this correlation may originate from a direct interaction between the gas and the dark matter (Qin \& Wu \cite{qin01}).

The correlation between $T_{\rm cut}$ and the ICM temperature and density
also suggests that the surrounding environment is also an important parameter to determine $T_{\rm cut}$.
We further speculate that a thermal coupling between the central region and the ICM has been taking place.
For example, a certain magnetic field configuration may transfer a large amount of energy from the ICM to the galactic cool gas as the member galaxies move through the ICM (Makishima et al \cite{makishima01} and references therein).
Alternatively, the jet from active galactic nuclei at the galaxy center and associated bubbles may stir the cool galactic gas into surrounding hot ICM gas (e.g. Churazov et al. \cite{churazov01}).
Note, however, that among six systems shown in Fig.~\ref{fig:comp} only M87 exhibits a well defined radio jet or bubble. 
NGC~5044 has only a compact radio core (Slee et al. \cite{slee94}).
One important test for this model is to check if this large scale mixing preserves the observed metallicity concentration around the cD and other giant galaxies (e.g. Fujita \& Kodama \cite{fujita95}; Fukazawa et al.\cite{fukazawa98}; Allen \& Fabian \cite{allen98}).

To investigate each model and understand the physics around central galaxies,
systematic observations of galaxies on different scales and in different environments
along with theoretical studies are crucial. 
For such attempts, we refer to Kaastra et al. (\cite{kaastra02b}) and Peterson et al. (\cite{peterson02}).

\subsection{Origins of metals in galaxies and intra-cluster medium}
Our accurate measurement of the thermal structure has 
in turn provided the determination of the Fe and O abundances,
more robustly than any previous measurements.
Using the EPIC data of NGC~5044, Buote et al. (\cite{buote02}) reported that the Fe abundance ($R<2'$) is $1\pm0.2$ relative to the {\em meteoritic} solar abundance\footnote{The meteoritic Fe abundance is $3.24\times10^{-5}$, which is 70\% of our {\em photospheric} value used.}.
This value corresponds to $\sim 0.7\pm0.15$~solar in our definition, 
and is consistent with our measurement, $0.55\pm0.05$, within the errors.

The Fe abundance of NGC~5044 we measured and that of NGC~4636 (Xu et al. \cite{xu01}), 0.5--1.0 times solar, 
are within the range of {\it ASCA} measurements in 27 ellipticals (Matsushita et al. \cite{matsushita00}).
Therefore we refer the reader to the discussion on the Fe abundances presented by Matsushita et al. (\cite{matsushita00}).
Most importantly, we have confirmed that the Fe abundance in the X-ray emitting gas is already comparable to that of the stellar component and consequently 
additional enrichment by supernovae is not substantially required, when no dilution by the in-flowing ICM is assumed.

Contrary to Fe, the determination of the O abundance in NGC~5044 is reported here for the first time.
The O/Fe ratio of NGC~5044, about half of the solar ratio,
is similar to that of NGC~4636 and other galaxies at the cluster core (e.g., Tamura et al. \cite{tamura01b}; Kaastra et al. \cite{kaastra01}),
indicating a common origin of the metals in these systems.
On the other hand, the O/Fe ratios in these ellipticals are about two times smaller than those in the intracluster medium (e.g. see the previous references) and the solar ratio.
This difference suggests the presence of at least two different origins for metals in the ellipticals and the ICM, 
regardless of any theoretical models.
Variations in the Si/Fe ratio within a cluster (e.g. Finoguenov et al. \cite{finoguenov00}) and among clusters (e.g. Fukazawa et al. \cite{fukazawa98}) also support a multiple origin of the metals in the ICM.
For example, Tamura et al. (\cite{tamura01b}) explained the change in the O/Fe ratio between the central galaxy and the ICM in A496, in terms of an increase of the relative contribution from type Ia supernovae to the metal production in the galaxy.
The present result supports this simple idea.

To further understand the metal production in clusters, 
systematic measurements of variations in the O/Fe ratio within a cluster and among clusters are important. 
Note that the O/Fe ratio is more sensitive to their origins than the Si/Fe ratio. 
For such a study the RGS has not enough sensitivity for spatially extended O lines. 
Future experiments such as X-ray microcalorimeter (XRS) on board Astro-E II would provide more accurate measurements. 
In addition to the O, Si, and Fe abundances, the C and N abundances are also valuable.
This is because C and N are believed to be produced largely by stellar mass loss instead of supernovae. 
To measure those abundances a higher sensitivity for lower energy X-ray lines (0.1--0.5~keV) is needed.

\begin{acknowledgements}
This work is based on observations obtained with {\it XMM-Newton}, an ESA science 
mission with instruments and contributions directly funded by 
ESA Member States and the USA (NASA).
SRON is supported financially by NWO, the Netherlands Organization for Scientific
Research. 
We thank J.~Peterson, T.~de Jong, J.~W.~A. den Herder, and the referee, J.~Vrtilek for useful comments.
\end{acknowledgements}

\end{document}